\documentclass[seceq]{ptptex}
\title{%        %You can use \\ for explicit line-break
Absence of coexisting phase of quark-antiquark and diquark 
condensed phases in the extended Gross-Neveu model in $2 + 1$ 
dimensions%
}

\author{%       %Use \scshape  for the family name
Akira \textsc{Ni\'{e}gawa}%
}

\inst{%     %Affiliation, neglected when [addenda] or [errata]
Graduate School of Science, Osaka City University, 
Sumiyoshi-ku, Osaka 558-8585, JAPAN
}

\abst{%         %this abstract is neglected when [addenda] or [errata]
We show that the coexisting phase of quark-antiquark and diquark 
condensed phases is absent in the cold quark matter in the $2 + 1$ 
dimensional extended Gross-Neveu model, which is in sharp contrast 
to the case of $3 + 1$ dimensional Nambu--Jona-Lasinio model. 
}

\begin{document}

\maketitle

\section{Introduction}
Extensive studies for last decade on quark matters have disclosed 
the existence of various phases. Among those are the quark-antiquark 
($q \bar{q}$) condensed phase \cite{hk}, different diquark ($q q$) 
condensed phases. Furthermore, it has been suggested \cite{sch} that 
the coexisting phase of $q \bar{q}$ and $q q$ condensations also 
exists. (For recent reviews, see, e.g., \cite{mei}.)

One of the important theoretical models for studying this issue is 
the extended Nambu--Jona-Lasinio (NJL) model, which is an low energy 
effective theory of QCD \cite{mei}. The Gross-Neveu (GN) model 
\cite{gn} proposed in 1974 is the counterpart of the NJL model in $1 
+ 1$ and $2 + 1$ dimensional spacetime. Study of the field theoretic 
models in lower spacetime dimension are interesting in itself as in 
the solid-state physics. Various works have been devoted to the 
study of the GN model. (References together with a brief survey of 
them are given in \cite{k}.) Recently, the phase structure of the $2 
+ 1$ dimensional (3D) {\em extended} GN model has been studied 
within the mean-field approximation \cite{k}. In \cite{k}, quarks 
are assigned to the lowest nontrivial (2-dimensional) representation 
of the $O (2, 1)$ group, which we refer to as the 2d-spinor quarks. 
The model contains essentially two parameters, the $q \bar{q}$ 
coupling constant $G_S$ and $q q$ coupling constant $G_D$. Through 
numerical analyses for some sets of values of $(G_S, G_D)$, it has 
been shown that there appears $q \bar{q}$- and $q q$-condensed 
phases but no coexisting phase appears, which is in sharp contrast 
to the case of NJL model. 

The purpose of this note is to analytically prove the absence of the 
coexisting phase for any values of $G_S$ $(> 0)$ and $G_D$. We 
restrict to the cold \lq\lq quark matter'' with zero temperature 
($T = 0$). 
%%%%%%%%%%%%%%%%%%%%%%%%%%%%%%%%%%%%%%%%%%%%%%%%%%%%%%%%%%%
%%%%%%%%%%%%%%%%%%%%%%%%%%%%%%%%%%%%%%%%%%%%%%%%%%%%%%%%%%%
%%%%%%%%%%%%%%%%%%%%%%%%%%%%%%%%
\setcounter{section}{1}
\setcounter{equation}{0}
\def\theequation{\mbox{\arabic{section}.\arabic{equation}}} 
%%%%%%%%%%%%%%%%%%%%%%%%%%%%%%%%%%%%%%%%%%%%%%%%%%%%%%%%%%%%
\section{Preliminary} 
The Lagrangian of the extended GN model with 2d-spinor quarks reads 
%%%%%%%%%%%%%%%%%%%%%%%%%%%%%%
\begin{equation}
{\cal L} = \bar{q} \left( i 
\partial\kern-0.em\raise0.17ex\llap{/}\kern0.15em\relax + \mu 
\gamma_0 \right) q + G_S (\bar{q} q)^2 + G_D \left( \bar{q} \tau_2 
\lambda_2 q^c \right) \left( \bar{q}^c \tau_2 \lambda_2 q \right) 
\, , 
\end{equation}
%%%%%%%%%%%%%%%%%%%%%%%%%%%%%%%%%%%%
where $\mu$ is the quark chemical potential. The quark fields, $q$ 
and $\bar{q}$, are the doublets in the flavor space and the triplets 
in the color space. $q^c$ and $\bar{q}^c$ are the charge-conjugated 
fields. The Pauli matrix $\tau_2$ acts on the flavor space, while 
the Gell-Mann matrix acts on the color space. Employing the 
mean-field approximation, we get 
%%%%%%%%%%%%%%%%%%%%%%%%%%%%%%
\begin{eqnarray}
{\cal L} & = & \bar{q} \left( i 
\partial\kern-0.em\raise0.17ex\llap{/}\kern0.15em\relax - \sigma + 
\mu \gamma_0 \right) q - \frac{1}{2} \Delta \left( \bar{q} \tau_2 
\lambda_2 q^c \right) - \frac{1}{2} \Delta^* \left( \bar{q}^c \tau_2 
\lambda_2 q \right) \nonumber \\ 
&& - \frac{\sigma^2}{4 G_S} - \frac{|\Delta|^2}{4 G_D} \, . 
\end{eqnarray}
%%%%%%%%%%%%%%%%%%%%%%%%%%%%%%%%%%%%
Here 
%%%%%%%%%%%%%%%%%%%%%%%%%%%%%%
\begin{equation}
\sigma = - 2 G_S \langle \bar{q} q \rangle \, , \;\;\;\; 
\Delta = - 2 G_D \langle \bar{q}^c \tau_2 \lambda_2 q \rangle \, , 
\end{equation}
%%%%%%%%%%%%%%%%%%%%%%%%%%%%%%%%%%%%
where $\sigma$ and $\Delta$ are the order parameters for the $q 
\bar{q}$ and $qq$ condensations, respectively. Computation of the 
thermodynamic potential at $T = 0$ yields \cite{k} 
%%%%%%%%%%%%%%%%%%%%%%%%%%%%%%
\begin{equation}
\Omega = \frac{\sigma^2}{4 G_S} + \frac{|\Delta|^2}{4 G_D} - 
\int_\sigma^\infty \frac{d E}{2 \pi} \, E \left[ 2 \left( E_\Delta^+ 
+ E_\Delta^- \right) + \left( E_\Delta^+ + E_\Delta^- 
\right)_{\Delta = 0} \right] \, , 
\label{Th-pot}
\end{equation}
%%%%%%%%%%%%%%%%%%%%%%%%%%%%%%%%%%%%
where 
%%%%%%%%%%%%%%%%%%%%%%%%%%%%%%
\begin{eqnarray}
\left( E_\Delta^\pm \right)^2 & = & E^2 + \mu^2 + |\Delta|^2 \pm 
2 \sqrt{E^2 \mu^2 + \sigma^2 |\Delta|^2} \, , \nonumber \\ 
E &=& \sqrt{\vec{p}^{\, 2} + \sigma^2} \, , \;\;\;\; E^\pm = E \pm 
\mu \, . 
\end{eqnarray}
%%%%%%%%%%%%%%%%%%%%%%%%%%%%%%%%%%%%
Changing the integration variable $E \to \xi$ through $\xi = 
\sqrt{E^2 + \sigma^2 |\Delta|^2 / \mu^2} \pm \mu$, the integrals in 
(\ref{Th-pot}) that contain $E_\Delta^\pm$ may be carried out 
analytically. We perform renormalization as in \cite{k}, and choose, 
without loss of generality, $\Delta$ to be real and positive. Then, 
we introduce following dimensionless quantities: 
%%%%%%%%%%%%%%%%%%%%%%%%%%%%%%
\begin{eqnarray}
x & = & \sigma / \mu \;\; (\geq 0) \, , \;\;\; y = \tilde{\Delta} / 
\mu \;\; (\geq 1) \, , \nonumber \\ 
A &=& - \frac{2 \pi}{\mu} \left( \frac{1}{4 G_S} - \frac{3 
\alpha}{8} \right) \, , \;\;\; B = - \frac{2 \pi}{\mu} \left( 
\frac{1}{4 G_D} - \frac{\alpha}{4} \right) \, , \nonumber \\ 
\Omega (x, y) &=& \frac{2 \pi}{\mu^3} \Omega (\sigma, 
\tilde{\Delta}) \, , 
\label{tikan}
\end{eqnarray}
%%%%%%%%%%%%%%%%%%%%%%%%%%%%%%%%%%%%
where $\alpha$ is the renormalization scale and $\tilde{\Delta} 
\equiv \sqrt{\Delta^2 + \mu^2}$. After all this, we obtain 
%%%%%%%%%%%%%%%%%%%%%%%%%%%%%%
\begin{eqnarray}
\Omega (x, y) & = & \frac{1}{3} \left\{ (x + 1)^3 + |x - 1|^3 
\right\} + \frac{1}{2} \left\{ (x - 1) |x - 1| - (x + 1)^2 \right\} 
\nonumber \\ 
&& + \frac{2}{3} \left\{ (x + y)^3 + |x - y|^3 \right\} + (x y - 1) 
|x - y| - ( x y + 1 ) ( x + y ) \nonumber \\ 
&& + \left( x^2 - 1 \right) \left( y^2 - 1 \right) \ln \frac{x y + 1 
+ x + y }{x y - 1 + |x - y|} \, . 
\label{potential}
\end{eqnarray}
%%%%%%%%%%%%%%%%%%%%%%%%%%%%%%%%%%%%
Note that, for the normal phase, $(x, y) = (0, 1)$, $\Omega (x = 0, 
y = 1) = -1$. Throughout in this paper, we restrict to the case 
$0 < A$, the case in which the quark-antiquark condensate can appear. 
%%%%%%%%%%%%%%%%%%%%%%%%%%%%%%%%%%%%
%% SUB %%%%%%%%%%%%%%%%%%%%%%%%%%%%%%%%%%
%%%%%%%%%%%%%%%%%%%%%%%%%%%%%%%%%%%%
\subsection{Quark-antiquark condensation} 
Here we study the case with $y = 1$ $(\Delta = 0)$: 
%%%%%%%%%%%%%%%%%%%%%%%%%%%%%%
\begin{equation}
\Omega_x (x) \equiv \Omega (x, y = 1) = (3 - A) x^2 - 1 \, . 
\end{equation}
%%%%%%%%%%%%%%%%%%%%%%%%%%%%%%%%%%%%
Straightforward analysis yields 
%%%%%%%%%%%%%%%%%%%%%%%%%%%%%%
\begin{equation}
\Omega_x \geq \left\{ 
\begin{array}{ll}
-1 & \;\;\; \mbox{for} \; A \leq 3 \\ 
- A^3 / 27 & \;\;\; \mbox{for} \; 3 \leq A 
\end{array}
\right\} \equiv \tilde{\Omega}_x 
\end{equation}
%%%%%%%%%%%%%%%%%%%%%%%%%%%%%%%%%%%%
%%%%%%%%%%%%%%%%%%%%%%%%%%%%%%%%%%%%
%% SUB %%%%%%%%%%%%%%%%%%%%%%%%%%%%%%%%%%
%%%%%%%%%%%%%%%%%%%%%%%%%%%%%%%%%%%%
\subsection{Diquark condensation} 
Here we study the case with $x = 0$ $(\sigma = 0)$: 
%%%%%%%%%%%%%%%%%%%%%%%%%%%%%%
\begin{eqnarray}
\Omega_y (y) & \equiv & \Omega (x = 0, y) \nonumber \\ 
& = & \frac{4}{3} y^3 - B (y^2 - 1) - 2 y - (y^2 - 1) L (y) - 
\frac{1}{3} \, , 
\end{eqnarray}
%%%%%%%%%%%%%%%%%%%%%%%%%%%%%%%%%%%%
where $L (y) \equiv \ln [ (y + 1) / (y - 1) ]$. $\Omega_y (y)$ is 
minimum at $y = y_0$: 
%%%%%%%%%%%%%%%%%%%%%%%%%%%%%%
\begin{eqnarray}
2 y_0 - L (y_0) & = & B \, , 
\label{jyouk} \\ 
\Omega_y (y_0) &=& - \frac{1}{3} \left( 2 y_0^3 + 1 \right) \, . 
\label{omegay}
\end{eqnarray}
%%%%%%%%%%%%%%%%%%%%%%%%%%%%%%%%%%%%
From (\ref{jyouk}), $\mbox{Max} (1, B / 2) < y_0$ and then we 
have, for any $B$, 
%%%%%%%%%%%%%%%%%%%%%%%%%%%%%%%%%%%%%%%%%%%%%%%%%%%
\begin{equation} 
\Omega_y (y_0) < - 1 = \Omega (x = 0, y = 1) \, . 
\label{fav}
\end{equation} 
%%%%%%%%%%%%%%%%%%%%%%%%%%%%%%%%%%%%%%%%%%%%%%%%%%%%
Thus, the diquark condensed state is energetically favored over the 
\lq\lq normal vacuum''. 

We now find the upper bound of $\Omega_y (y)$ in the region $2 \leq 
B$, which we will use extensively in the sequel. Since $L (y)$ 
monotonically decreases with increasing $y$ and its curvature is 
positive, 
%%%%%%%%%%%%%%%%%%%%%%%%%%%%%%
\begin{eqnarray}
L (y) & \leq & L (B / 2) + \left[ \frac{d L (y)}{d y} \right]_{y = B 
/ 2} \left( y - \frac{B}{2} \right) \nonumber \\ 
&=& - \frac{8}{B^2 - 4} y + \frac{4 B}{B^2 - 4} + L (B / 2) \, . 
\end{eqnarray}
%%%%%%%%%%%%%%%%%%%%%%%%%%%%%%%%%%%%
Substituting this into (\ref{jyouk}), we have 
%%%%%%%%%%%%%%%%%%%%%%%%%%%%%%
\begin{equation}
B \leq 2 y_0 + \frac{8}{B^2 - 4} y_0 - \frac{4 B}{B^2 - 4} - L (B / 
2) \, . 
\end{equation}
%%%%%%%%%%%%%%%%%%%%%%%%%%%%%%%%%%%%
Solving this, we obtain 
%%%%%%%%%%%%%%%%%%%%%%%%%%%%%%
\begin{equation}
y_0 \geq \frac{B^2 - 4}{2 B} + \frac{2}{B} + \frac{B^2 - 4}{2 
B^2} \, L (B / 2) \, . 
\label{rt}
\end{equation}
%%%%%%%%%%%%%%%%%%%%%%%%%%%%%%%%%%%%%%%
Here and in the following, we frequently use the inequality: 
%%%%%%%%%%%%%%%%%%%%%%%%%%%%%%%%%%%%%%%%%%%%%%%%%
\begin{equation}
2 (y - 1) \leq (y^2 - 1) L (y) \leq 2 y \;\;\;\;\;\; (1 \leq y) \, . 
\label{ineq1} 
\end{equation}
%%%%%%%%%%%%%%%%%%%%%%%%%%%%%%%%%%%%%%%%%%%%%%%%%%%%%%%%%%%%
Using this in (\ref{rt}), we obtain 
%%%%%%%%%%%%%%%%%%%%%%%%%%%%%%%%%%%%%%%%%%%%
\begin{equation}
y_0 \geq \frac{B}{2} + \frac{4 (B - 2)}{B^2} \;\;\; \left( \equiv 
y_0' \right) \, . 
\label{iie} 
\end{equation}
%%%%%%%%%%%%%%%%%%%%%%%%%%%%%%%%%%%%

Substituting (\ref{iie}) into (\ref{omegay}), we obtain 
%%%%%%%%%%%%%%%%%%%%%%%%%%%%%%
\begin{equation}
\Omega_y (y_0) \leq - \frac{1}{3} \left( 2 y_0^{' \, 3} + 1 \right) 
\leq - \frac{B^3}{12} - B + \frac{5}{3} \;\;\; \left( \equiv 
\tilde{\Omega}_y \right) \;\;\;\;\;\;\; (2 \leq B) \, . 
\end{equation}
%%%%%%%%%%%%%%%%%%%%%%%%%%%%%%%%%%%%

We will use $\tilde{\Omega}_y$, an the upper bound of $\Omega_y$. 
Numerical computation shows that $\left[ \Omega_y (y_0) \right]_{B 
= 2} = - 3.515...$, while $\left[ \tilde{\Omega}_y \right]_{B = 2} 
= -1$. 

In the following sections we show that $\Omega \geq \mbox{Min} 
\left( \tilde{\Omega}_x, \, \tilde{\Omega}_y \right)$ $>$ 
$\mbox{Min} \left( \tilde{\Omega}_x, \, \Omega_y \right)$ for 
any value of $(0 <) A$ and $B$, so that the quark-antiquark and 
diquark condensates do not coexist . 
%%%%%%%%%%%%%%%%%%%%%%%%%%%%%%%%%%%%%%%%%%%%%%
%%%% SEC III %%%%%%%%%%%%%%%%%%%%%%%%%%%%%%%
%%%%%%%%%%%%%%%%%%%%%%%%%%%%%%%%%%%%%%%%%%%%%%
\setcounter{section}{2}
\setcounter{equation}{0}
\def\theequation{\mbox{\arabic{section}.\arabic{equation}}} 
%%%%%%%%%%%%%%%%%%%%%%%%%%%%%%%%%%%%%%%%%%%%%%%%%%%%%%%%%%%%
\section{Region I ($x \leq 1 \leq y$)} 
%%%%%%%%%%%%%%%%%%%%%%%%%%%%%%%%%%%%%%%%%%%%%%
%%%% SUBSEC %%%%%%%%%%%%%%%%%%%%%%%%%%%%%%%
%%%%%%%%%%%%%%%%%%%%%%%%%%%%%%%%%%%%%%%%%%%%%%
We divide the whole region of $x$ and $y$ into the three regions; 
$(0 \leq ) \, x \leq 1 \leq y$ (Region I), $(1 \leq) \, y \leq x$ 
(Region II) and $1 < x < y$ (Region III). In this section, we 
analyze $\Omega (x, y)$ in the Region I. Analyses in the Region II 
and Region III will be made in subsequent sections. 

In the region $( 0 \leq ) \, x \leq 1 \leq y$ (Region I), $\Omega 
(x, y)$ reads 
%%%%%%%%%%%%%%%%%%%%%%%%%%%%%%
\begin{eqnarray}
\Omega (x, y) & = & F (y) x^2 + \frac{4}{3} y^3 - B (y^2 - 1) - 2 y 
- (y^2 - 1) L (y) - \frac{1}{3} \, , 
\label{potential1} \\ 
F (y) & = & 1 + 2 y - A + (y^2 - 1) L (y) \, . 
\label{efu}
\end{eqnarray}
%%%%%%%%%%%%%%%%%%%%%%%%%%%%%%%%%%%%
For $0 \leq F (y)$, $\Omega (x, y) \geq \Omega (x = 0, y) = 
\Omega_y (y)$. Then we restrict to the region $F (y) \leq 0$, where 
%%%%%%%%%%%%%%%%%%%%%%%%%%%%%%
\begin{eqnarray}
\Omega (x, y) & \geq & \Omega (x = 1, y) \nonumber \\ 
&=& \frac{4}{3} y^3 - B (y^2 - 1) - A + \frac{2}{3} \;\;\; \left( 
\equiv \tilde{\Omega}_I (y) \right) \, . 
\label{hosi}
\end{eqnarray}
%%%%%%%%%%%%%%%%%%%%%%%%%%%%%%%%%%%%
$\tilde{\Omega}_I (y)$ is minimum at $y = B / 2$ and monotonically 
increases with increasing $(B / 2 <) \, y$. Since $1 \leq y$ for $B 
\leq 2$, $\tilde{\Omega}_I (y) \geq \tilde{\Omega}_I (y = 1) = 2 - A 
\geq \tilde{\Omega}_x$. The equality holds for $A = 3$, 
$\tilde{\Omega}_I = - 1$. Since $\Omega_y (y_0) < - 1$ (Eq. 
(\ref{fav})), $\tilde{\Omega}_I > \Omega_y (y_0)$ there. 

For $2 \leq B$, $\Omega (x, y) \geq \tilde{\Omega}_I (y = B / 2)$. 
We first study 
%%%%%%%%%%%%%%%%%%%%%%%%%%%%%%%%%%%%%%%%%%%%%%%%%%
\begin{equation}
\Delta_y \tilde{\Omega}_I (y = B / 2) \equiv \tilde{\Omega}_I ( y = 
B / 2) - \tilde{\Omega}_y = 2 B - A - 1 \, , 
\end{equation}
%%%%%%%%%%%%%%%%%%%%%%%%%%%%%%%%%%%%%%%%%%%%%%%%%%%%%%%%%
which is nonpositive for 
%%%%%%%%%%%%%%%%%%%%%%%%%%%%%%%%%%%%%%%%%%%%%%%%%%
\begin{equation}
(3 \leq) \;\; 2 B - 1 \leq A \;\;\; \mbox{or} \;\;\; B \leq 
\frac{A + 1}{2} \, . 
\label{arya} 
\end{equation}
%%%%%%%%%%%%%%%%%%%%%%%%%%%%%%%%%%%%%%%%%%%%%%%%%%%%%%%%%
We next study 
%%%%%%%%%%%%%%%%%%%%%%%%%%%%%%%%%%%%%%%%%%%%%%%%%%
\begin{equation}
\Delta_x \tilde{\Omega}_I (y = B / 2) \equiv \tilde{\Omega}_I (y = B 
/ 2) - \tilde{\Omega}_x = \frac{A^3}{27} - \frac{B^3}{12} + B - A + 
\frac{2}{3} 
\end{equation}
%%%%%%%%%%%%%%%%%%%%%%%%%%%%%%%%%%%%%%%%%%%%%%%%%%%%%%%%%
in the region (\ref{arya}). $\Delta_x \tilde{\Omega}_I (y = B / 2)$ 
monotonically decreases with increasing $B$ (for $2 < B$). Then, from 
(\ref{arya}), $\Delta_x \tilde{\Omega}_I (y = B / 2) \geq \left[ 
\Delta_x \tilde{\Omega}_I (y = B / 2) \right]_{B = (A + 1) / 2}$ 
$(\equiv H (A))$, which is a polynomial in $A$. It can easily be 
seen that $H (A)$ increases with increasing $A$ ($3 \leq A$) and 
$0 \leq H (A)$. Equality holds for $A = 3$ and then $B = 2$, where 
$H (A) > \Omega_y (y_0)$. 

Note that we have not used the constraint on $y$ that comes from 
$F (y) \leq 0$ (see above after (\ref{efu})).  
%%%%%%%%%%%%%%%%%%%%%%%%%%%%%%%%%%%%%%%%%%%%%%%%%%%%%%%%%%%%%%%%%%%
%%% SEC IV %%%%%%%%%%%%%%%%%%%%%%%%%%%%%%%%%%%%%%%%%%%%%%%%%%%%%%%%%%
%%%%%%%%%%%%%%%%%%%%%%%%%%%%%%%%%%%%
%%%%%%%%%%%%%%%%%%%%%%%%%%%%%%%%%%%%%%%%%%%%%%%%%%%%%%%%%%%%
\section{Region II ($(1 \leq) \, y \leq x$)} 
In this region, 
%%%%%%%%%%%%%%%%%%%%%%%%%%%%%%
\begin{eqnarray}
\Omega (x, y) & = & 2 x^3 - A x^2 + (y^2 - 1) G (x) 
\label{potential3} \\ 
G (x) & = & 2 x - B + (x^2 - 1) L (x) \, . 
\label{jii}
\end{eqnarray}
%%%%%%%%%%%%%%%%%%%%%%%%%%%%%%%%%%%%
For $0 \leq G (x)$, $\Omega (x, y) \geq \Omega (x, y = 1) = 
\Omega_x (x)$, and then we restrict to the region $G (x) \leq 0$, 
where 
%%%%%%%%%%%%%%%%%%%%%%%%%%%%%%
\begin{eqnarray}
\Omega (x, y) & \geq & \Omega (x, y = x) \nonumber \\ 
&=& 4 x^3 - (A + B) x^2 - 2 x + (x^2 - 1)^2 L (x) + B \;\;\; \left( 
\equiv \tilde{\Omega}_{II} (x) \right) \, . 
\label{hosi1}
\end{eqnarray}
%%%%%%%%%%%%%%%%%%%%%%%%%%%%%%%%%%%%
$\tilde{\Omega}_{II} (x)$ is minimum at $x = x_0$: 
%%%%%%%%%%%%%%%%%%%%%%%%%%%%%%%%%%%%%%%%%%%%%%%%%%%%%
\begin{eqnarray}
&& 5 x_0 + 2 (x^2_0 - 1) L (x_0) = A + B \, , 
\label{jyouB1} \\ 
&& \tilde{\Omega}_{II} (x_0) = \frac{1}{2} \left[ 3 x_0^2 - (A + B) 
x_0^2 + x_0 + B - A \right] \, . 
\label{tilome1} 
\end{eqnarray}
%%%%%%%%%%%%%%%%%%%%%%%%%%%%%%%%%%%%%%%%%%%%%%%%%%%%%
$(1 <)$ $x_0$ exists for $5 \leq A + B$. Using (\ref{ineq1}) in 
(\ref{jyouB1}), we obtain 
%%%%%%%%%%%%%%%%%%%%%%%%%%%%%%%%%%%%
\begin{equation}
x_0 \leq (A + B + 4) / 9 \, . 
\label{4.3d}
\end{equation}
%%%%%%%%%%%%%%%%%%%%%%%%%%%%%%%%%%%%%%%%%
%%%%%%%%%%%%%%%%%%%%%%%%%%%%%%%%%%%%%%%%%%%%%%%%%%%%%
%%%%%%%%%%%%%%%%%%%%%%%%%%%%%%%%%%%%%%%%%%%%%%%%%%%%%
%%%%%%%%%%%%%%%%%%%%%%%%%%%%%%%%%%%%%%%%%%%%%%%%%%%%%
\subsubsection*{The region $A + B \leq 5$}
>From (\ref{4.3d}), for $A + B \leq 5$, $x_0 \leq 1$ and then 
$\tilde{\Omega}_{II} (x)$ monotonically increases with increasing 
$(1 \leq) \, x$, so that $\tilde{\Omega}_{II} (x) \geq 
\tilde{\Omega}_{II} (x = 1) = 2 - A \geq \tilde{\Omega}_x$. 
%%%%%%%%%%%%%%%%%%%%%%%%%%%%%%%%%%%%%%%%%%%%%%%%%%%%%
%%%%%%%%%%%%%%%%%%%%%%%%%%%%%%%%%%%%%%%%%%%%%%%%%%%%%
%%%%%%%%%%%%%%%%%%%%%%%%%%%%%%%%%%%%%%%%%%%%%%%%%%%%%
\subsubsection*{The region $5 \leq A + B$}
With the help of (\ref{ineq1}), the condition $G (x) \leq 0$ (see 
above after (\ref{jii})) yields $B \geq 2 x + (x^2 - 1) L (x) \geq 4 
x - 2$, from which we have 
%%%%%%%%%%%%%%%%%%%%%%%%%%%%%%%%%%%%%%%%%%%%%%%%%%%%%
\begin{equation}
x \leq \frac{B + 2}{4} \, . 
\label{jyouB2}
\end{equation}
%%%%%%%%%%%%%%%%%%%%%%%%%%%%%%%%%%%%%%%%%%%%%%%%%%%%%
Since $1 \leq x$, $2 \leq B$. 

In the following, putting aside the equation (\ref{jyouB1}), we 
regard $\tilde{\Omega}_{II} (x_0)$ in (\ref{tilome1}) as a function 
of $x_0$ and show that $\Delta \tilde{\Omega}_{II} (x_0) > 0$. It can 
be shown that, in the region of our interest, $1 \leq x_0 \leq (A + 
B + 4) / 9$, the polynomial $\tilde{\Omega}_{II} (x_0)$ 
monotonically decreases with increasing $x_0$. 

We recall the condition, Eq. (\ref{jyouB2}). 

\underline{$(A + B + 4) / 9 \leq (B + 2) / 4$ or $A \leq (5 B + 2) / 
4$}: In this region, $\tilde{\Omega}_{II} (x_0) \geq 
\tilde{\Omega}_{II} (x_0 = (A + B + 4) / 9)$, which is a polynomial 
in $A$ and $B$. $\Delta_y \tilde{\Omega}_{II} (x_0 = (A + B + 4) / 
9)$ $\, (\equiv \Delta_y \hat{\Omega} (A, B))$ is a polynomial in 
$A$ and $B$. It can be shown that, in the region of our interest, 
this function decreases monotonically in $A$. Since $A \leq (5 B + 2) 
/ 4$, $\Delta_y \hat{\Omega} (A, B) \geq \Delta_y \hat{\Omega} (A = 
(5 B + 2) / 4, B)$, which is a polynomial in $B$. This polynomial 
increases monotonically with increasing $(2 \leq) \, B$ and is 
positive. After all of this, we have $\Delta \tilde{\Omega}_{II} 
(x, y) > 0$. 

\underline{$(B + 2) / 4 < (A + B +4) / 9$ or $(5 B + 2) / 4 < A$}: 
In this region, $\tilde{\Omega}_{II} (x_0) \geq \tilde{\Omega}_{II} 
(x_0 = (B + 2) / 4)$. The polynomial $\Delta_x \tilde{\Omega}_{II} 
(x_0 = (B + 2) / 4) \, (\equiv \Delta_x \hat{\Omega} (A, B))$ 
increases monotonically with increasing $A$ (in the region of our 
interest). Since $(5 B + 2) / 4 \leq A$, $\Delta_x \hat{\Omega} (A, 
B) \geq \Delta_x \hat{\Omega} (A = (5 B + 2) / 4, B)$, which is a 
polynomial in $B$. This polynomial increases monotonically with 
increasing $(2 \leq) \, B$ and $\Delta_x \hat{\Omega} (A = (5 B + 2) 
/ 4, B) \geq 0$. The equality holds for $B = 2$ and then $A = 3$. At 
$(A, B) = (3, 2)$, $\hat{\Omega} - \Omega_y > 0$. 
%%%%%%%%%%%%%%%%%%%%%%%%%%%%%%%%%%%%%%%%%%%%%%%%%%%%%%%%%%%%%%%%%%%
%%% SEC V %%%%%%%%%%%%%%%%%%%%%%%%%%%%%%%%%%%%%%%%%%%%%%%%%%%%%%%%%%
%%%%%%%%%%%%%%%%%%%%%%%%%%%%%%%%%%%%
%%%%%%%%%%%%%%%%%%%%%%%%%%%%%%%%%%%%%%%%%%%%%%%%%%%%%%%%%%%%
\section{Region III ($1 < x < y$)} 
In this region, 
%%%%%%%%%%%%%%%%%%%%%%%%%%%%%%
\begin{eqnarray}
\Omega (x, y) & = & \frac{2}{3} x^3 + \left( 2 y - A + (y^2 - 
1) L (y) \right) x^2 \nonumber \\ 
&& + \frac{4}{3} y^3 - 2 y - B (y^2 - 1) - (y^2 - 1) L (y) \, . 
\label{potential2}
\end{eqnarray}
%%%%%%%%%%%%%%%%%%%%%%%%%%%%%%%%%%%%
With respect to $x$, $\Omega (x, y)$ in (\ref{potential2}) is 
minimum at $x = x_0$: 
%%%%%%%%%%%%%%%%%%%%%%%%%%%%%%%%%%%%%%%%%%%%%%%%%%%%%
\begin{equation}
x_0 = A - 2 y - (y^2 - 1) L (y) \, . 
\label{iiya}
\end{equation}
%%%%%%%%%%%%%%%%%%%%%%%%%%%%%%%%%%%%%%%%%%%%%%%%%%%%%%%%%%%%%%%%%%%
The following four regions should be studied: 

Region IIIA: (\ref{iiya}) does not have a solution in the region, 
$1\leq x_0$ and $1 \leq y$. 

Region IIIB: $x_0 \leq 1$. 

Region IIIC: $(1 \leq) \, y \leq x_0$. 

region IIID: $1 < x_0 < y$. 
%%%%%%%%%%%%%%%%%%%%%%%%%%%%%%%%%%%%%%%%%%%%%%%%%%%%%%%%%%%%%
%% SUB %%%%%%%%%%%%%%%%%%%%%%%%%%%%%%%%%%%%%%%%%%%%%%%%%%%%%%
%%%%%%%%%%%%%%%%%%%%%
\subsection*{Region IIIA and Region IIIB} 
$\Omega (x, y)$ in (\ref{potential2}) monotonically increases with 
increasing $(1 \leq) \, x$. Then, $\Omega (x, y) \geq \Omega (x = 1, 
y)$, which is the same as (\ref{hosi}) in the case of Region I 
above. Then, $\Delta \Omega (x, y) > 0$. 
%%%%%%%%%%%%%%%%%%%%%%%%%%%%%%%%%%%%%%%%%%%%%%%%%%%%%%%%%%%%%
%% SUB %%%%%%%%%%%%%%%%%%%%%%%%%%%%%%%%%%%%%%%%%%%%%%%%%%%%%%
%%%%%%%%%%%%%%%%%%%%%
\subsection*{Region IIIC} 
$\Omega (x, y)$ monotonically decreases with increasing $x$ ($1 < x 
\leq y$). Then, $\Omega (x, y) \geq \Omega (x = y, y)$, which is the 
same as $\tilde{\Omega}_{II} (y)$ in (\ref{hosi1}) in the case of 
Region II above, and then we briefly describe. In the present case, 
from (\ref{iiya}) with (\ref{ineq1}), the condition $y \leq x_0$ 
yields 
%%%%%%%%%%%%%%%%%%%%%%%%%%%%%%%%%%%%%%%%%%%%%%%%%%%%%%%%%%%%%%
\begin{equation}
( 1 \leq ) y = x \leq \frac{A + 2}{5} \, , 
\label{condD1}
\end{equation}
%%%%%%%%%%%%%%%%%%%%%%%%%%%%%%%%%%%%%%%%%%%%%%%%%%%%%%%%%
which is the counterpart of (\ref{jyouB2}). From (\ref{condD1}) we 
have $A \leq 3$. We recall here (\ref{4.3d}). 

\underline{$(A + B + 4) / 9 \leq (A + 2) / 5$ or $B \leq 2 (2 A - 1) 
/ 5$}: In this region, $\tilde{\Omega}_{II} (x_0)$ in (\ref{tilome1}) 
satisfies $\tilde{\Omega}_{II} (x_0) \geq \tilde{\Omega}_{II} (x_0 = 
(A + B + 4) / 9)$, and the function $\tilde{\Omega}_{II} (x_0 = (A + 
B + 4) / 9) - \tilde{\Omega}_x \, (\equiv \Delta_x \hat{\Omega} (A, 
B))$ is a polynomial in $A$ and $B$. It can be shown that, in the 
region of our interest, this function decreases monotonically in $B$. 
Since $B \leq 2 (2 A - 1)/5$, $\Delta_x \hat{\Omega} (A, B) \geq 
\Delta_x \hat{\Omega} (A, B = 2 (2 A - 1)/5)$, which is polynomial 
in $A$. This polynomial increases monotonically with increasing $(3 
\leq) \, A$ and $\Delta_x \hat{\Omega} (A, B =  2 (2 A - 1)/5) \geq 
0$. The equality holds at $A = 3$ and then $B = 2$, where 
$\hat{\Omega} - \Omega_y > 0$. 

\underline{$(A + 2) / 5 \leq (A + B +4) / 9$ or $2 (2 A - 1) / 5 < 
B$}: In this region, $\Delta_y \tilde{\Omega}_{II} (x_0) \geq 
\Delta_y \tilde{\Omega}_{II} (x_0 = (A + 2) / 5)$ $(\equiv \Delta_y 
\hat{\Omega} (A, B))$, which increases monotonically with increasing 
$B$. Since $2 (2 A - 1) / 5 < B$, $\Delta_y \hat{\Omega} (A, B) \geq 
\Delta_y \hat{\Omega} (A, B = 2 (2 A - 1) / 5)$, which is a 
polynomial in $A$. This polynomial increases with increasing $(3 
\leq) \, A$ and is positive. 
%%%%%%%%%%%%%%%%%%%%%%%%%%%%%%%%%%%%
%% SUBSEC %%%%%%%%%%%%%%%%%%%%%%%%%%%%%%%%%%%%%%%%%%%%%%%%%%%%%%%%
%%%%%%%%%%%%%%%%%%%%%%%%%%%%%%%%%%%%%%%%%%%%%%%%%%%%%%%%%%%%
\subsection{Region IIID} 
In this region, $\Omega (x, y) \geq \Omega ( x_0, y)$ (Eq. 
(\ref{potential2})): 
%%%%%%%%%%%%%%%%%%%%%%%%%%%%%%
\begin{equation}
\Omega ( x_0, y) = - \frac{x_0^2}{3} + x_0 + \frac{4}{3} y^3 - B 
(y^2 - 1) - A \, , 
\label{omega1}
\end{equation}
%%%%%%%%%%%%%%%%%%%%%%%%%%%%%%%%%%%%
where $x_0$ is as in (\ref{iiya}). 

Using (\ref{ineq1}), the condition $1 < x_0 < y$ yield 
%%%%%%%%%%%%%%%%%%%%%%%%%%%%%%%%%%%%%%%%%%%%
\begin{equation}
\mbox{Max} \left( 1, \frac{A}{5} \right) < y < \frac{A + 1}{4} \, , 
\label{sibari}
\end{equation}
%%%%%%%%%%%%%%%%%%%%%%%%%%%%%%%%%%%%%%%%%%%%%%%%%%%%%%%%%%%%
from which we can restrict to the region $A \leq 3$ because $1 \leq 
y$. 

Ignoring the explicit form of $x_0$, we analyze the function $\Omega 
(x_0, y)$. Since $\Omega (x_0, y)$ monotonically decreases with 
increasing  $(1 \leq) \, x_0$, 
%%%%%%%%%%%%%%%%%%%%%%%%%%%%%%%%%%%%%%%%%%%%
\begin{eqnarray}
\Omega (x_0, y) & > & \Omega (x_0 = y, y) = y^3 - B (y - 1) + y - A 
\nonumber \\ 
& \equiv & \tilde{\Omega} (A, B; y) \, . 
\end{eqnarray}
%%%%%%%%%%%%%%%%%%%%%%%%%%%%%%%%%%%%%%%%%%%%%%%%%%%%%%%%%%%%

For $B \leq 2$, $\tilde{\Omega}$ monotonically increases with 
increasing $(1 \leq ) y$, and then $\tilde{\Omega} (A, B; y) \geq 
\tilde{\Omega} (A, B; y = 1) = 2 - A \geq \tilde{\Omega}_x$. 

For $2 \leq B$, $\tilde{\Omega} (A, B; y)$ is minimum at 
%%%%%%%%%%%%%%%%%%%%%%%%%%%%%%%%%%%%%%%%%%%%
\begin{eqnarray}
y_0 & = & \frac{1}{3} \left( B + \sqrt{B^2 - 3} \right) \, , 
\nonumber \\ 
\tilde{\Omega} (A, B; y_0) & = & - \frac{2}{27} (B^2 - 3) \left( 
B + \sqrt{B^2 - 3} \right) + \frac{10}{9} B - A \, . 
\end{eqnarray}
%%%%%%%%%%%%%%%%%%%%%%%%%%%%%%%%%%%%%%%%%%%%%%%%%%%%%%%%%%%%

For $y < y_0$ $(y_0 < y)$, $\tilde{\Omega} (A, B; y)$ monotonically 
decreases (increases) with increasing $y$. From (\ref{sibari}), $y 
\leq (A + 1)/4$. Solving $y_0 (B = B_0) = (A + 1) / 4$ for $B$, we 
obtain 
%%%%%%%%%%%%%%%%%%%%%%%%%%%%%%%%%%%%%%%%%%%%
\begin{equation}
B_0 (A) = \frac{1}{8 (A + 1)} \left[ 3 (A + 1)^2 + 3 (A + 1) + 
16 \right] \, , 
\end{equation}
%%%%%%%%%%%%%%%%%%%%%%%%%%%%%%%%%%%%%%%%%%%%%%%%%%%%%%%%%%%%
which is an increasing function of $(3 <) A$. Then, 
%%%%%%%%%%%%%%%%%%%%%%%%%%%%
\begin{equation}
\left\{ 
\begin{array}{ll}
\mbox{For} \, B \leq B_0 (A) \, , \;\;\; & y_0 \leq (A + 1) / 4 
\, , \\  
\mbox{For} \, B_0 (A) < B \, , \;\;\; & (A + 1) / 4 < y_0 \, , 
\end{array}
\right. 
\end{equation}
%%%%%%%%%%%%%%%%%%%%%%%%%%%%%%
%% SUBSUB %%%%%%%%%%%%%%%%%%%%%%%%%%%%
%%%%%%%%%%%%%%%%%%%%%%%%%%%%%%
\subsubsection*{Region: $\, 2 \leq B \leq B_0 (A)$ $(3 \leq A)$} 
We first analyze $\Delta_x \tilde{\Omega} \geq A^3/27 + 
\tilde{\Omega} (A, B_0; y_0 (B_0))$. Numerical computation shows 
that $A^3/27 + \tilde{\Omega} (A, B_0; y_0 (B_0)) \leq 0$ for $A < 
4.173...$. Then, we analyze $\Delta_y \tilde{\Omega}$ for $A \leq 
4.18$: 
%%%%%%%%%%%%%%%%%%%%%%%%%%%%%%%%%%%%%%%%%%%%
\begin{eqnarray}
\Delta_y \tilde{\Omega} (A, B; y_0) & \geq & \Delta_y \tilde{\Omega} 
(A = 4.18, B; y_0) \nonumber \\ 
& = & \frac{B^3}{12} + \frac{19}{9} B - \frac{4}{27} B (B^2 - 3) - 
\frac{5}{3} - 4.18 \, . 
\end{eqnarray}
%%%%%%%%%%%%%%%%%%%%%%%%%%%%%%%%%%%%%%%%%%%%%%%%%%%%%%%%%%%%
It can easily be seen that this polynomial is positive for $A \leq 3 
\leq 4.18$. 
%%%%%%%%%%%%%%%%%%%%%%%%%%%%%%
%% SUBSUB %%%%%%%%%%%%%%%%%%%%%%%%%%%%
%%%%%%%%%%%%%%%%%%%%%%%%%%%%%%
\subsubsection*{Region: $\, B_0 (A) \leq B$ $(3 \leq A)$} 
In this case, 
%%%%%%%%%%%%%%%%%%%%%%%%%%%%%%%%%%%%%%%%%%%%
\begin{eqnarray}
\tilde{\Omega} (A, B; y) & \geq & \tilde{\Omega} (A, B; y = (A + 1) 
/ 4) \nonumber \\ 
& = & \frac{(A + 1)^3}{64} - \frac{3}{4} (A + 1) + 1 - \frac{1}{16} 
\left[ (A + 1)^2 - 16 \right] B \, . 
\end{eqnarray}
%%%%%%%%%%%%%%%%%%%%%%%%%%%%%%%%%%%%%%%%%%%%%%%%%%%%%%%%%%%%
>From this, we see that $\Delta_x \tilde{\Omega} \leq 0$ for 
%%%%%%%%%%%%%%%%%%%%%%%%%%%%%%%%%%%%%%%%%%%%
\begin{eqnarray}
B & \geq & \frac{1}{108} \left( 91 (A + 1) - 192 + \frac{352}{A + 5} 
\right) \nonumber \\ 
& \geq & \frac{1}{108} [ 91 (A + 1) - 192 ] \;\;\; \left( \equiv B_1 
(A) \right) \, . 
\end{eqnarray}
%%%%%%%%%%%%%%%%%%%%%%%%%%%%%%%%%%%%%%%%%%%%%%%%%%%%%%%%%%%%

Numerical computation shows that, for $A \leq A_0 = 4.396...$ $\; [ 
A_0 < A ]$, $B_1 \leq B_0$ $\; [ B_0 < B_1 ]$. Then, the region of 
our interest is as follows: 
%%%%%%%%%%%%%%%%%%%%%%%%%%%%
\begin{equation}
\left\{ 
\begin{array}{ll}
\mbox{For} \, A \leq A_0 \, , \;\;\; & B_0 (A) \leq B 
\, , \\  
\mbox{For} \, A_0 < A \, , \;\;\; & B_1 (A) < B \, . 
\end{array}
\right. 
\end{equation}
%%%%%%%%%%%%%%%%%%%%%%%%%%%%%%

Now, we analyze $\Delta_y \tilde{\Omega}$ for $B_1 (A) \leq B$ and 
$B_0 (A) \leq B$, which is polynomial in $A$ and $B$. For $A + 1 
\leq \sqrt{32}$, this polynomial monotonically increases with 
increasing $B$, while, for $\sqrt{32} < A + 1$, this polynomial 
takes minimum at $B = B_2 (A)$: 
%%%%%%%%%%%%%%%%%%%%%%%%%%%%%%%%%%%%%%%%%%%%%%%%%%%%%%%%%%%%
\begin{equation}
B_2 (A) = \frac{1}{2} \sqrt{(A + 1)^2 - 32} \, . 
\end{equation}
%%%%%%%%%%%%%%%%%%%%%%%%%%%%%%%%%%%%%%%%%%%%%%%%%%%%%%%%%%%%%%%%%%%%%

\underline{$\sqrt{32} < A + 1$}: Numerical computation shows that, 
%%%%%%%%%%%%%%%%%%%%%%%%%%%%
\begin{equation}
\left\{ 
\begin{array}{ll}
\mbox{For} \, A \leq A_2 = 9.84... \, , \;\;\; & B_2 \leq B_0 \, , 
\\ 
\mbox{For} \, A_2 < A \, , \;\;\; & B_0 < B_2 \, . 
\end{array}
\right. 
\end{equation}
%%%%%%%%%%%%%%%%%%%%%%%%%%%%%%
On the other hand, one can analytically show that $B_2 < B_1$. After 
all of this, we learn that the region (of $A$ and $B$) that should be 
analyzed is $(A_0, B_1) \leq (A, B)$. In this region, $\Delta_y 
\tilde{\Omega} (A, B) > \Delta_y \tilde{\Omega} (A, B_1)$, which is 
polynomial in $A$. It is straightforward to show that this 
polynomial $\Delta_y \tilde{\Omega} (A, B_1)$ is positive for $A_0 
\leq A$. 

\underline{$A + 1 < \sqrt{32}$}: From the above analysis, we see 
that, for $A_0 \leq A \leq \sqrt{32} - 1$, $B_1 \leq B$, and, for 
$(3 \leq) A < A_0$, $B_0 < B$. In the former case, $\Delta_y 
\tilde{\Omega}$ is polynomial in $A$ and is positive. In the latter 
case, 
%%%%%%%%%%%%%%%%%%%%%%%%%%%%%%%%%%%%%%%%%%%%%%%%%%%%%%%%%%%%
\begin{eqnarray}
\Delta_y \tilde{\Omega} & \geq & \Delta_y \tilde{\Omega} (B = B_0) 
\nonumber \\ 
& \geq & \Delta_y \tilde{\Omega} (B = B_0') \, , 
\label{qp}
\end{eqnarray}
%%%%%%%%%%%%%%%%%%%%%%%%%%%%%%%%%%%%%%%%%%%%%%%%%%%%%%%%%%%%%%%%%%%%%
where 
%%%%%%%%%%%%%%%%%%%%%%%%%%%%%%%%%%%%%%%%%%%%%%%%%%%%%%%%%%%%
\begin{equation}
B_0' (A) = \frac{1}{8} \left( 3 (A + 1) + 3 + \frac{16}{4.4 + 1} 
\right) \;\; \left( < B_0 (A) \right) \, , 
\end{equation}
%%%%%%%%%%%%%%%%%%%%%%%%%%%%%%%%%%%%%%%%%%%%%%%%%%%%%%%%%%%%%%%%%%%%%
which holds in the region under study. One can see that he last 
expression in (\ref{qp}), which is polynomial in $A$, is positive in 
the region of interest. 
%%%%%%%%%%%%%%%%%%%%%%%%%%%%%%%%%%%%%%%%%%%%%%%%%%%%%%%%%%%%%%%%%%%%%
%%%%%%%%%%%% SEC VI %%%%%%%%%%%%%%%%%%%%%%%%%%%%%%%%%%%%%%%%%%%%%%%%%%%%
%%%%%%%%%%%%%%%%%%%%%%
\section{Concluding remark} 
We have studied the phase structure of the extended 3D Gross-Neveu 
model with 2d-spinor quarks. The following results are obtained. 
\begin{description}
\item{(I)} There does not appear the region where the 
quark-antiquark and diquark condensations coexist. 
\item{(II)} The diquark condensed state is always energetically 
favored over the normal vacuum (see after (\ref{fav})). 
\item{(III)} From (I) and (II), we see that, for a given $q \bar{q}$ 
coupling constant $A$, there is the critical value, $B_c$ for the 
$q q$ coupling constant $B$: For $B \leq B_c$, $q \bar{q}$-condensed 
state is realized, while, for $B_c < B$, the $q q$-condensed state 
is realized. 
\end{description}
%%%%%%%%%%%%%%%%%%%%%%%%%%%%%%%%%%%%%%%%%%%%%%%%%%%%%%%%%%%%%%%%%%
%%%%%%%%%%%%%%%%%%%%%%%%%%%%%%%%%%%%%%%%%%%%%%%%%%%%%%%%%%%%%%%%%%
%%%%%%%%%%%%%%
\section*{Acknowledgement}
I would like to thank M. Inui and H. Kohyama for useful discussions.

%\section*{Acknowledgements}
%We would like to thank ...........

%\appendix
%\section{First Appendix} %Empty argument \section{} yields `Appendix'. 
%
%\section{Second Appendix}

\end{document}